\documentclass[%
 reprint,longbibliography,preprintnumbers,
nofootinbib,
 amsmath,amssymb,
 aps
]{revtex4-1}
\pdfoutput=1
\usepackage{graphicx}
\usepackage[utf8]{inputenc}
\usepackage{flushend}
\usepackage{dcolumn}
\usepackage{bm}
\usepackage{balance}

\usepackage[normalem]{ulem}

\usepackage[colorlinks = true,
            linkcolor = blue,
            urlcolor  = blue,
            citecolor =green,
            anchorcolor = blue]{hyperref}
\usepackage{verbatim}
\usepackage{color,ulem}
\usepackage[english]{babel}

\usepackage[utf8]{inputenc}
\input Starburst.fd
\newcommand*\initfamily{\usefont{U}{Starburst}{xl}{n}}\initfamily

\newcommand{\beq}{\begin{eqnarray}}
\newcommand{\eeq}{\end{eqnarray}}
\usepackage{amsmath}
\usepackage{tikz}
\usetikzlibrary{decorations.pathmorphing}
\usetikzlibrary{shapes.misc}
\tikzset{cross/.style={cross out, draw=black, minimum size=8*(#1-\pgflinewidth), inner sep=0pt, outer sep=0pt},
cross/.default={1pt}}
\usetikzlibrary{patterns,math}
\begin{document}

\title{\Large Reply to W. T. Kranz on ``Explicit analytical solution for random close packing in $d=2$ and $d=3$''}

\author{\textbf{Alessio Zaccone}$^{1,2}$}%
 \email{alessio.zaccone@unimi.it}
 
 \vspace{1cm}
 
\affiliation{$^{1}$Department of Physics ``A. Pontremoli'', University of Milan, via Celoria 16,
20133 Milan, Italy.}
\affiliation{$^{2}$Cavendish Laboratory, University of Cambridge, JJ Thomson
Avenue, CB30HE Cambridge, U.K.}


\maketitle
In a response to our paper \cite{mypaper}, W. Till Kranz (WTK) \cite{KranzArXiv}, present some erroneous considerations, that are in contradiction with all the previous literature, and which can be summarized as follows.
WTK, first of all, claims that our paper ``contains mathematical flaws'', which  is not true since our original paper \cite{mypaper} is (i) mathematically correct (as anyone can easily verify) under the stated assumptions and (ii) all the assumptions are critically and carefully discussed in the paper.

WTK claims that ``As the contact value $g(\sigma)$ of the pair correlation function
$g(r)$ diverges at close packing [2], Eq. (6) in Ref. [1] appears suspicious.''
It is of course true that the $g(r)$ diverges at contact, this is well known from simulations, but this is not in contradiction with Eq. (6) in our paper \cite{mypaper}.
As a matter of fact Eqs. (5)-(6) in our paper are fully consistent with Eq. (17) in \cite{Torquato_2018} used by Torquato to effectively represent the divergence of the $g(r)$ at contact, i.e. by means of a Dirac delta function which takes care of the \emph{discrete} part of the probability distribution function. 

After this, WKT introduces his own description of the short-range radial distribution function $g(r)$ of the RCP, as follows:
\begin{equation}
g(r) = g_{sing}(r/\sigma) + g_{reg}(r/\sigma),
\end{equation}
i.e. by splitting the $g(r)$ into a singular (diverging) part and into a regular (non-diverging) part. WTK's choice is at odds not only with Eq. (5) in our paper, but, even more importantly, is inconsistent with Eq. (17) in the recent review by Torquato \cite{Torquato_2018} which summarizes the state of the art in the field.
In particular, we argue that the above equation proposed by WKT is incorrect because the correct splitting is not between ``singular'' and ``regular'', which has no meaning in terms of theory of probability distribution functions (pdfs), but rather between ``discrete'' and ``continuous'', as correctly argued in our paper \cite{mypaper} and by Torquato \cite{Torquato_2018}.
In the theory of pdfs the splitting between discrete and continuous parts is standard and discussed in most treatises on probability theory.
WTK's splitting of the $g(r)$ into singular and regular parts has no meaning, and is clearly erroneous, in terms of probability theory, because the $g(r)$ is a pdf and thus cannot be singular (because obviously, as is well known, it has to be normalized).

This erroneous assumption in WKT's Comment \cite{KranzArXiv} invalidates the author's further results, which are based on it, and the claims about our paper.



\begin{figure}[ht]
    \includegraphics[width=0.45\textwidth]{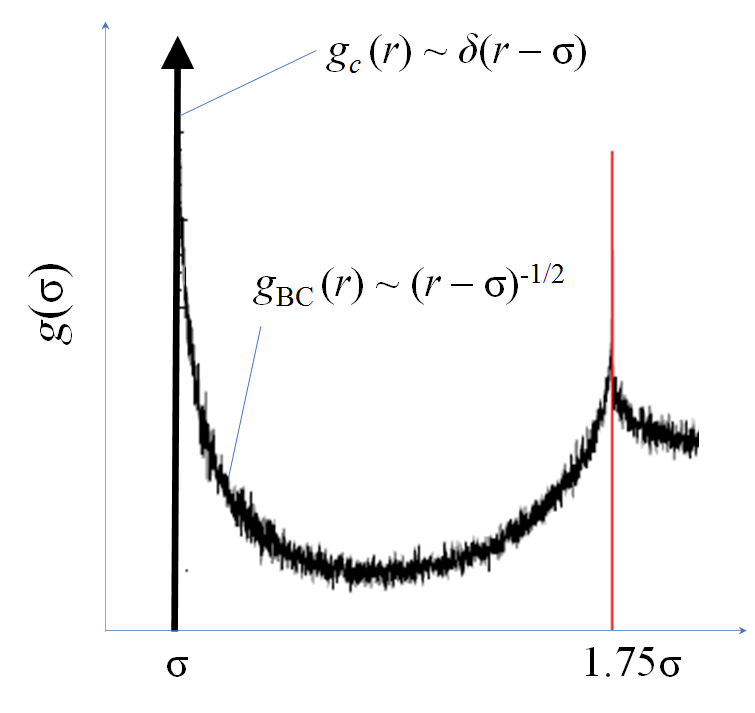}
    \caption{The near-contact radial distribution function $g(r)$ at random close packing in $d=3$. The black curve represents numerical data adapted from \cite{Donev}, this is the \emph{continuous} part of the $g(r)$, according to \cite{mypaper,Torquato_2018} (note that a slightly different exponent $0.4$ was discovered in \cite{Lerner} and appears in agreement with replica theory \cite{Zamponi}). The thick vertical arrow represents the Dirac delta in the \emph{discrete} part of $g(r)$ according to Eqs.(5)-(6) of Ref. \cite{mypaper} and Eq. (17) of Ref. \cite{Torquato_2018}. This picture is also consistent with all the previous literature on the $g(r)$ of random aggregates of spheres, e.g. \cite{Weitz} and \cite{Morbidelli}. }
    \label{fig:my_label}
\end{figure}

In summary, the correct picture for the RCP's $g(r)$ as a pdf, from both the point of view of probability theory and physical considerations, is schematized in Fig. 1. This picture is consistent with both the definition used in our paper \cite{mypaper}, and that commonly accepted in the random packing literature \cite{Torquato_2018}, as well as with a number of previous works on the mathematical modelling of the near-contact part of the $g(r)$ of random aggregates of spheres, e.g. \cite{Weitz} and \cite{Morbidelli}. 

\bibliographystyle{apsrev4-1}

\end{document}